\documentclass[
final,3p,twocolumn,times
]{elsarticle}
\usepackage[utf8]{inputenc}
\usepackage[T1]{fontenc}
\usepackage{listings}

\usepackage{xcolor}

\definecolor{codegreen}{rgb}{0,0.6,0}
\definecolor{codegray}{rgb}{0.5,0.5,0.5}
\definecolor{codepurple}{rgb}{0.58,0,0.82}
\definecolor{backcolour}{rgb}{0.95,0.95,0.92}

\lstdefinestyle{mystyle}{
    backgroundcolor=\color{backcolour},   
    commentstyle=\color{codegreen},
    keywordstyle=\color{magenta},
    numberstyle=\tiny\color{codegray},
    stringstyle=\color{codepurple},
    basicstyle=\ttfamily\footnotesize,
    breakatwhitespace=false,         
    breaklines=true,                 
    captionpos=b,                    
    keepspaces=true,                 
    numbers=left,                    
    numbersep=5pt,                  
    showspaces=false,                
    showstringspaces=false,
    showtabs=false,                  
    tabsize=2
}

\newcommand\lib[1]{\texttt{#1}}

\makeatletter
\def\ps@pprintTitle{%
  \let\@oddhead\@empty
  \let\@evenhead\@empty
  \let\@oddfoot\@empty
  \let\@evenfoot\@oddfoot
}
\makeatother

\usepackage[hidelinks]{hyperref}
\begin{document}

\begin{frontmatter}

\title{OSSCAR, an open platform for collaborative development of computational tools for education in science}

\author[mymainaddress,secondaddress,CECAM]{Dou Du}
\author[CECAM]{Taylor J. Baird}
\author[CECAM]{Sara Bonella}
\author[mymainaddress,secondaddress]{Giovanni Pizzi\corref{mycorrespondingauthor}}
\cortext[mycorrespondingauthor]{Corresponding author}
\ead{giovanni.pizzi@epfl.ch}

\address[mymainaddress]{Theory and Simulation of Materials (THEOS), École Polytechnique Fédérale de Lausanne, CH-1015 Lausanne, Switzerland}

\address[secondaddress]{National Centre for Computational Design and Discovery
of Novel Materials (MARVEL), École Polytechnique Fédérale de Lausanne, CH-1015 Lausanne, Switzerland}

\address[CECAM]{CECAM Centre Européen de Calcul Atomique et Moléculaire, École Polytechnique Fédérale de Lausanne, CH-1015 Lausanne, Switzerland}

\begin{abstract}
In this paper we present the Open Software Services for Classrooms and Research (OSSCAR) platform. 
OSSCAR provides an open collaborative environment to develop, deploy and access educational resources in the form of web applications.
To minimize efforts in the creation and use of new educational material, it combines software tools that have emerged as standards with custom domain-specific ones.
The technical solutions adopted to create and distribute content are described and motivated on the basis of reliability, sustainability, ease of uptake and use.
Examples from courses in the domains of physics, chemistry, and materials science are shown to demonstrate the style and level of interactivity of typical applications.
The tools presented are easy to use, and create a uniform and open environment exploitable by a large community of teachers, students, and researchers with the goal of facilitating learning and avoiding, when possible, duplication of efforts in creating teaching material.
Contributions to expand the educational content of the OSSCAR project are welcome.
\end{abstract}


\end{frontmatter}

%\linenumbers

\section{\label{sec:level1} Introduction}

Software-based tools, such as notebooks or illustrative codes, are increasingly
employed in scientific courses to enrich and complement more standard teaching
approaches.  These tools can provide an interactive environment for teachers to
demonstrate, via live examples and engaging visualization, complex and abstract
concepts that may otherwise be difficult to transmit.  At the same time,
students can gain intuition, facilitate understanding and strengthen
learning~\cite{deJong2013} by exploiting them as simple virtual laboratories,
e.g., to experiment in real time with the effect of relevant parameters in
equations.
Given these advantages, and with the growing relevance of remote education, software-based educational tools are becoming more common. For example, Quantum Physics Online~\cite{QuantumPhysicsOnline} publishes online Java applets with visualizations that illustrate topics typically covered in undergraduate and master level courses in that area. Considering more domain-specific examples, the Soft Matter Demos~\cite{softmatterdemos} or NanoHub~\cite{nanohub} websites present simulations and visualizations to stimulate interest in these domains, with a limited interest in coursework. Other open-source platforms offer visual tools for education in chemistry in the form of interactive simulations~\cite{LabXchange}. Use of e-tools based on Google Colab was recently explored~\cite{Vallejo2022} to support the teaching of thermodynamics and provide some introduction to coding in chemistry classes taught in Columbia. Further impetus to develop on-line teaching tools was added by the recent pandemic crisis~\cite{ijerph18052672,youmans2020,Zalat2021}, with several interesting studies on their effectiveness~\cite{10.3389/feduc.2021.638470,PhysRevPhysEducRes.17.020111,doi:10.1021/acs.jchemed.1c00655}.

In spite of their great potential, widespread adoption and
sharing of software-based tools for teaching is, however, still hindered by different barriers.  On the side of
the teachers, the time investment to create bespoke material for different
classes might be considerable and efforts frustrated by the lack of agile
development and deployment environments.
Moreover, curating the material to counteract software obsolescence, guaranteeing resilience to changes in versioning of the adopted language, and facilitating updates when content evolves are all non-trivial challenges.
Furthermore, given that the same type
of material is needed for classes across different areas and in different
institutions, the risk of effort duplication is very high.
No ``public library'' of software teaching tools exists to reduce this risk and
limit the teachers' effort only to the creation of new, original material.  On
the side of the students, uptake and usage of these tools can be problematic
depending on the technology employed to deploy them and the level of
user-friendliness of the platforms to access them.  Also, the lack of a coherent
platform may force them to spend considerable effort to migrate from one
technological solution to another when changing class.

In this paper, a new platform that attempts to overcome these barriers is
presented: the Open Software Services for Classrooms and Research (OSSCAR).
OSSCAR is inspired by the software architecture developed as part of
AiiDAlab~\cite{Yakutovich2021}, a platform designed to provide easy access to
research-oriented software, workflows and tools via a web interface.  OSSCAR
adapts and applies these concepts and technologies for education purposes.
Specifically, as detailed in Sec.~\ref{sec:developing-apps}, OSSCAR combines and
builds upon a set of well-established software tools to create a web-based
collaborative environment targeted at providing educational resources and
enhancing awareness and adoption of best practices in Open Science.  The
programming language chosen is Python, and we use
\lib{Jupyter}~\cite{jupyter,JupyterURL,Granger2021} and
\lib{JupyterLab}~\cite{JupyterLabURL} as the programming interface and
environment.  Within this framework, common visualization tools and widgets are
employed (and new ones are developed) to create interactive notebooks
illustrating specific topics or proposing exercises. \lib{Jupyter} notebooks are
then automatically converted into web applications exploiting the \lib{Voila}
program~\cite{VoilaURL}.  The web applications hide all code and show only the
outputs (including, in particular, the widgets to interact with the application)
in a user-friendly format accessible through any browser, circumventing the need
for specific software installation and set up.

The contributions of OSSCAR are along three main lines, detailed in the rest of
the paper, and that we summarize here: 1) provide custom graphical components
(widgets) for domain-specific visualization types (see
Sec.~\ref{sec:widgets}); 2) provide custom educational content tailored
for a number of courses in the fields of computational physics, chemistry and
materials science; these are developed in the form of self-contained modules
that can be combined and reused also beyond the courses for which they
were originally developed (see Sec.~\ref{sec:example-apps}); 3) provide clear
documentation (on \url{http://www.osscar.org}), transferring know-how gathered
in the past years on how to combine various open technologies to easily develop
new educational applications (see Sec.~\ref{sec:docs}) and open to input and
feedback from the community.

In the following, we first demonstrate the appearance and structure of an OSSCAR
notebook via the example of a classic problem in quantum mechanics:
the double-well potential.  We then provide technical details on the tools
employed to develop new notebooks and justify our choices.  Next, we show
some selected examples, based on notebooks developed for Master-level courses in
Physics, Chemistry and Materials Science. These examples, all related to
quantum-mechanical problems, also suggest that OSSCAR notebooks can act as
modules to be combined in various ways for different classes.

While technologically mature, OSSCAR is at the early stages of development in
terms of content.  It is intended as an open repository to be built in
collaboration with the community of students and teachers in scientific
disciplines, and we invite contributions from interested groups.

\section{\label{sec:firstexample}An OSSCAR interactive web application example: a quantum-mechanical double-well potential}

\begin{figure*}[tb]
   \centering\includegraphics[width=1.0\linewidth]{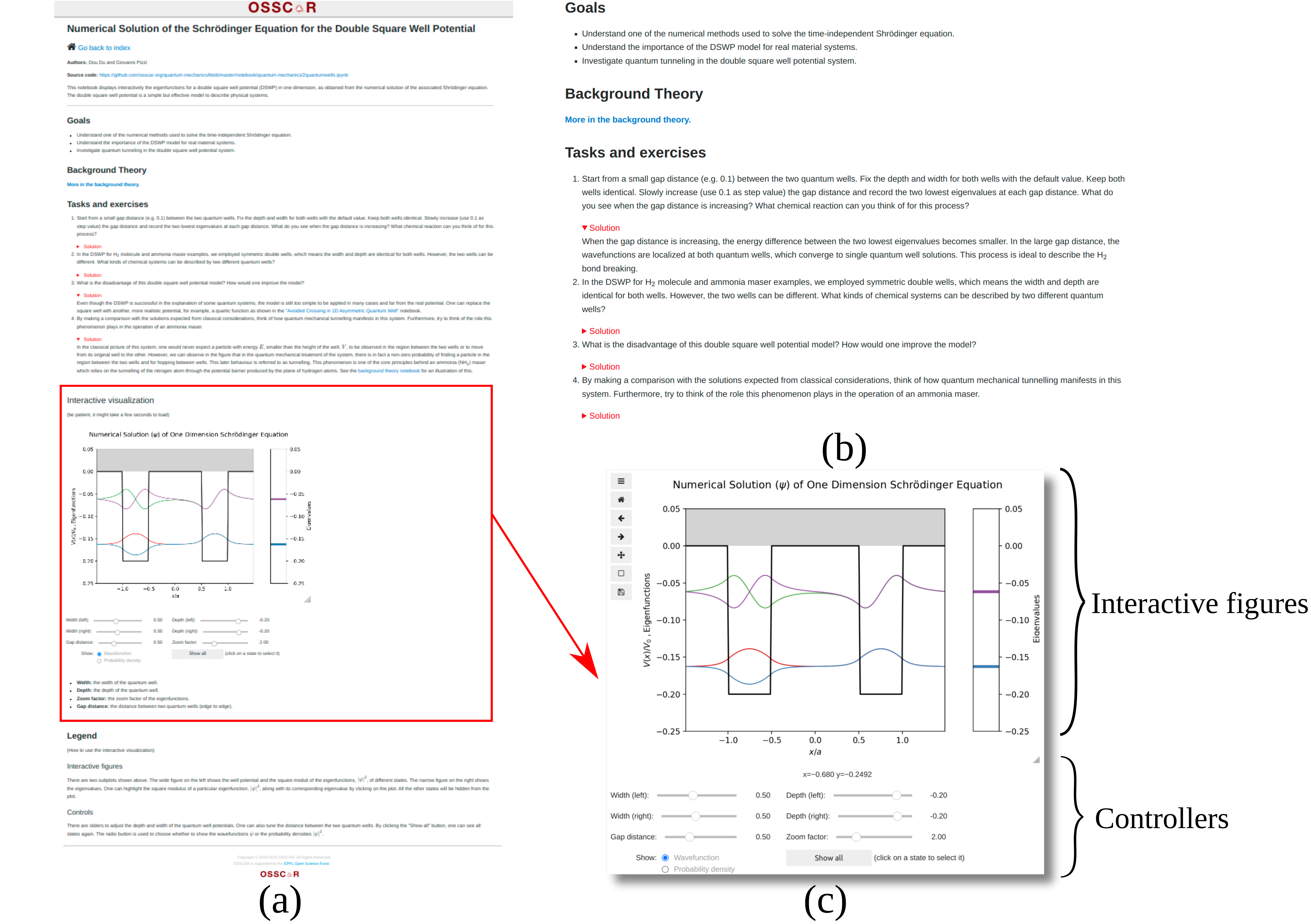}
    \caption{\label{fig:2quantumwells} The interactive web application to
        demonstrate the numerical solution of the Schrödinger equation for a 1D
        double square-well potential.  The web application is hosted at
        \url{https://osscar-quantum-mechanics.materialscloud.io}.  a) The whole
        webpage, showing the typical structure of an OSSCAR notebook:
        introductory text (zoomed in panel b) including a link to the background
        theory, the goals of the notebook and a set of tasks and exercises for
        the students, followed by the interactive visualization (better
        displayed in panel c), and a legend.  }
\end{figure*}

In this section, we discuss a prototypical OSSCAR application: the interactive
visualization of the eigenvalues and eigenvectors obtained by solving the
one-dimensional (1D) Schrödinger equation for a double square-well potential
\cite{Grosso2013}. To provide a general overview of a typical application, we mainly
focus on the key components that are shared among the OSSCAR notebooks, rather
than on the specific content of this example. Note that, since the web
applications are implemented as \lib{Jupyter} notebooks, as mentioned in the
Introduction and described in more detail in Sec.~\ref{sec:developing-apps}, in
the following we will use the terms notebooks and applications (almost)
interchangeably. 

To clarify the reasons behind the overall structure and graphical appearance of
the application, we start by discussing how we expect these notebooks to be
employed by teachers and students. In our experience, there are two main use
cases, that we shall call A and B. In use case
A, the notebooks are used for independent learning by students. Students might
have simply found the application online, or might have been referred to it with
a web link, for example in a class. In this case, before they are presented with
the interactive visualization itself, it is important to provide users with
a short explanation of the goals of the application and with some guidance on
how to best interact with it to achieve the learning objectives.

In use case B, instead, teachers might use the applications for live
demonstrations during their classes to complement and enrich standard lecturing
(textbooks, blackboard, slides, \ldots) thus improving students' learning
effectiveness~\cite{Hake1998}. We include in use case B also the case of
teaching assistants that present the notebook content as part of their
discussion sessions, because the requirements are relatively similar.  In this
second scenario, the introductory part of a notebook is not relevant, since the
topic has already been introduced and discussed by the teachers, who will
instead focus on using the interactive part. In this case, it is essential that
applications can be accessed very rapidly (in a matter of seconds), because they
will be used only for the minimal time required to convey the message (typically
no more than a couple of minutes), before the teachers continue with their
lectures. (The students might also use the applications after the lecture is
ended to revise the course content, falling back into use case A.)

With these two use cases in mind, we now discuss the structure of a typical
OSSCAR notebook, as illustrated in Fig.~\ref{fig:2quantumwells}.  Each
application starts with a brief textual section that we see at the top of
Fig.~\ref{fig:2quantumwells}(a), including a short list of educational goals, a
link to additional background theory, as well as a list of tasks to guide the
exploration of the interactive part of the notebook. Authorship is also
acknowledged at the top of each notebook to give due credit to contributors and
encourage collaborative contributions by other teachers and also by students.
This first section addresses the needs of use case A, to quickly assess if
the notebook covers the topics of interest and to provide guidance for interacting with the application via a set of tasks for the students.  At the
same time, this section is kept to a minimum to cover the needs of use case B
(or of students already familiar with the application): being short, the
section is easy to skip, so one can jump directly to the interactive
visualization.  In particular, the background theory is discussed in a
different, linked, page (and also there only as a brief overview of the physical
    problem, favoring links to existing online material to avoid content
duplication).  Furthermore, the solutions for the students' tasks are hidden by
default.  This latter design choice not only keeps the first textual section
short, but also encourages students to answer the questions themselves rather
than read directly the solutions, promoting active learning and thus improving
learning effectiveness~\cite{Crouch2004,Freeman2014}. 

Below this textual introduction, we find the interactive visualization section,
better displayed in Fig.~\ref{fig:2quantumwells}(b). This is the core of the web application.
Each interactive visualization is composed of two main groups of components: the
interactive figures and the controllers.  The controllers are ``widgets''
(discussed in more detail in Sec.~\ref{sec:widgets}) such as sliders, dropdown
menus, checkboxes or buttons, that allow one to tune some parameters of the model or
the visualization itself and whose effect is dynamically reflected in the
interactive figures.

In this specific example, the top part with the figures displays the potential
energy (thick black line, formed by two square wells) and the wavefunctions
(colored thin lines) at the height of the corresponding eigenvalues.  The right
part of the plot displays the eigenvalues only, represented as thick horizontal
lines (with the same color of the corresponding wavefunctions). 

In the controllers section of this figure, five sliders are used to tune the
width and depth of the two square wells and the distance between the two.  Two
radio buttons allow the students to decide whether to display the wavefunction
$\psi(x)$ or the probability density $|\psi(x)|^2$.  A sixth slider can be used
to determine the ``zoom factor'' of the wavefunctions (i.e., the multiplicative
factor in front of the wavefunction, that is only used to have a nice
visualization but does not affect the simulation).

In typical notebooks the figures, in addition to being dynamic (i.e., changing
their content as soon as the value of one of the controllers is modified), can
also be interactive: for instance, in this specific example, a click on one of
the wavefunctions (or on the corresponding eigenvalue on the right-hand side)
displays its plot and numerical value, while hiding all other wavefunctions.  A
button ``Show all'' in the controllers section allows one to display again all the
wavefunctions.

Finally, at the bottom of the page (see Fig.~\ref{fig:2quantumwells}(a)), there
is a legend that describes in more detail the figure components and the
functionality of each controller.  This is placed at the bottom of the page as a
useful reference, mostly to cover the needs of use case A, but we strive to
design the interactive visualizations so that all figures and controls are as
intuitive and self-explanatory as possible, reducing to a minimum the need to
read the legend.

A collection of OSSCAR web applications can be considered as a ``living book'',
with powerful interaction and visualization capabilities that go beyond what is
achievable on printed text or static images, and can convey more effectively
advanced content to students, facilitating their understanding.  In addition,
the tasks presented at the top of the notebook help students to
focus their attention on core concepts.  For instance, one of the tasks of
this notebook suggests investigating the phenomenon of quantum tunneling and
anticrossing of states as a function of the distance between the two wells: by
moving the slider to alter the gap distance, students can vividly observe in
real time how the wavefunctions and their energies change, something that would
be difficult to achieve through traditional teaching.
  
Finally, at the very top of each page, we also provide a link to the source code
of the notebook. We discuss the additional advantages of providing immediate
access to the notebook source code in Sec.~\ref{sec:docs}.

\section{\label{sec:developing-apps} Technology to develop interactive web applications}
One of our key design goals for OSSCAR is to make it simple enough for teachers
with basic coding experience to develop further applications.  Consequently, as
mentioned in the Introduction, the majority of the software stack is deliberately
composed of existing open, well documented and well maintained software. The
motivation behind doing this is to maximize the lifetime and accessibility of
the notebooks by ensuring that they do not depend on custom software and
technology that might become unsupported soon. 
Let us now detail the core technological components of the OSSCAR platform.

\subsection{Development environment: Python and \lib{Jupyter}/\lib{JupyterLab}}
We choose Python as the programming language for the interactive notebooks.
Python is a common programming language for data science and scientific
computing that has gained popularity in the past years in many computational
scientific disciplines~\cite{Perkel2015}.  This is probably due to both Python's
syntax, which is relatively easy to learn and quite readable even for people with little
programming experience, and to the very large number of free Python packages
that can be easily installed via, e.g., the \lib{pip}~\cite{Pip} or
\lib{conda}~\cite{Conda} package management tools.

Python allows for relatively rapid development, even if (being an interpreted
language) it might be slow for expensive computations.  The need for performant
simulations is less of an issue for education-oriented applications than for
scientific production runs, since the main goal is not to obtain results with
ultimate precision and speed, but rather to demonstrate the simplest approximation that
captures the essential aspects of the model (so that students can focus on the
core concepts, and not on the numerical optimizations). In spite of this, in
both use cases described in Sec.~\ref{sec:firstexample} it is very important
that simulations can be performed almost in real time (or in any case, in a
matter of seconds). Indeed, in use case A students might easily lose focus if
they have to click a button and wait for minutes before the results appear.
Moreover, slow execution time limits the interactive capabilities of the
applications and the number of different input parameters that students can
experiment with. Similarly, teachers in use case B need to be able to
demonstrate rapidly the relevant results to students before continuing with
their lectures.

While in our experience Python is often fast enough, there are cases
in OSSCAR where strategies to speed up the simulations are required (e.g., when
performing simulations with millions of iterations, or when dealing with large
matrices). We list some of these strategies in~\ref{app:speedup}.

For the purpose of creating interactive visualizations, however, the programming
language itself is not sufficient, but one also needs a library to enable
powerful displays of the results via a graphical user interface (GUI).
A relatively large number of GUI libraries are available for Python.  Our choice
is to use \lib{Jupyter} notebooks (using a ``classic'' \lib{Jupyter} server, or
the more recent \lib{JupyterLab} environment), that provide a notebook interface
to interact with Python code\footnote{We note for completeness that
\lib{Jupyter(Lab)} can actually work also with programming languages other than
Python.}. \lib{Jupyter} has very rapidly gained popularity, also in the
scientific context~\cite{Perkel2018,Barba2019} (including for teaching~\cite{Weiss2017}), as a very powerful approach to distribute
understandable and reusable code.

Having a notebook interface means that the whole code is divided in cells, and
each cell contains only a part of the code that can be executed independently,
and whose output is displayed underneath the cell.  The advantage is that the
notebooks do not contain only the source code, but can also include contextual
rich-text annotations, descriptions, and widgets (such as plots, buttons, \ldots
-- see discussion in Sec.~\ref{sec:widgets}), combining in a single consistent
document both the code and its documentation and visualization.

One of the reasons why we choose \lib{Jupyter} in OSSCAR, besides its popularity
(and thus availability of visualization and widget libraries), is that the GUI
is web-based and displays directly in the web browser.  As such, it works on any
computer operating system (OS) and does not require additional installation of
custom software, contrary to typical GUIs that might require one to install
OS-specific code and libraries.  Having a simple web-based interface is one of
the key requirements in OSSCAR to make the use of the applications
straightforward.  We discuss in the next sections how to implement interactive
visualizations within \lib{Jupyter} notebooks, and then discuss in
Sec.~\ref{sec:voila} how to completely hide the Python code and provide students
with a very simple and intuitive web interface.

\subsection{\label{sec:widgets}Widgets: components for interaction}
Visualization plays a crucial role in human learning~\cite{Gilbert2008}. This is
important, for instance, when dealing with multi-dimensional representations,
where interactive 3D plots can be very effective in representing datasets.
In
particular, in fields such as computational physics, interactive figures can
greatly facilitate explanation of abstract concepts compared to text, bare
equations or static figures.
In \ref{app:2D3D-plotting} we discuss some plotting libraries for 2D and 3D plots that we use in OSSCAR, with some minimal usage examples.

However, while plots are essential to display the results of a simulation or the
values of a function, one additional type of component is crucial to enable
interactivity and decide the relevant parameters in the controller section
(e.g., the width or depth of the quantum well in the example of
Fig.~\ref{fig:2quantumwells}).
As previously mentioned, these components are called widgets and allow the user
both to supply inputs and to trigger events (e.g., via a click on a button).
The Python library \lib{ipywidgets} provides a large number of common native widgets working within \lib{Jupyter} notebooks, such as sliders, checkboxes,
dropdowns, text areas and buttons.

We show in \ref{app:widgets-example} a simple example, both to illustrate how easy the code to interact with these widgets can be, and to discuss how one can implement instantaneous reaction to events.

\subsubsection{\label{sec:custom-widgets}OSSCAR custom widgets}
\begin{figure}[tb]
   \centering \includegraphics[width=7.5cm]{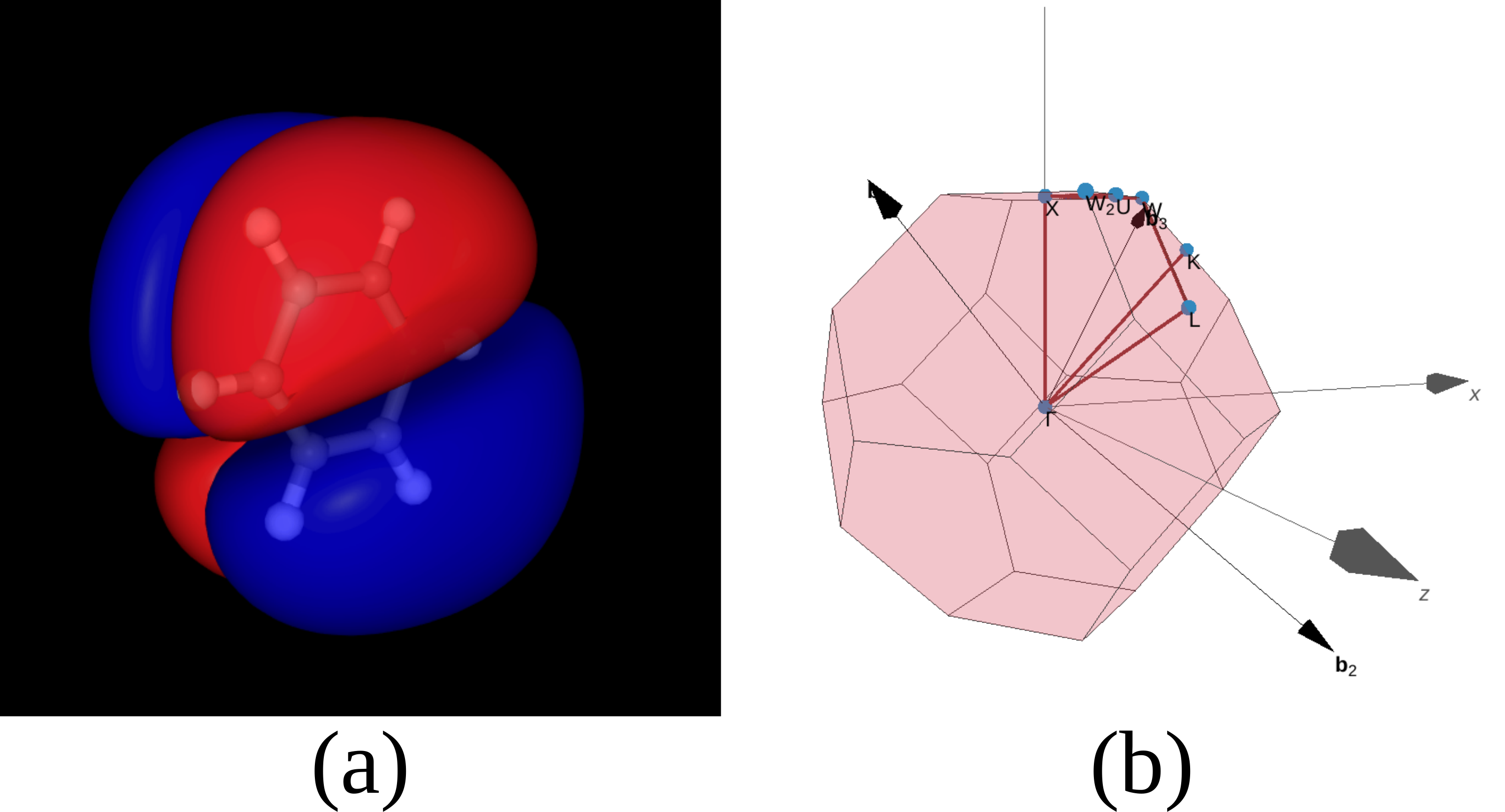}
    \caption{\label{fig:widgets} (a) A custom OSSCAR widget to display the
    molecular orbitals of a molecule (in this example, benzene), wrapping the
NGLView \lib{Jupyter} widget. (b) A custom OSSCAR widget to compute and plot the
first BZ of any crystal, together with the labels of high-symmetry points and a
suggested path to compute a band structure.} 
\end{figure}

When \lib{Jupyter}-ready widgets are not already available, OSSCAR develops
bespoke widgets customized for specific needs, embedded in our applications and
realized as open source. A typical example are custom visualizations necessary
to display domain-specific content, especially for 3D visualization.  While
general-purpose 3D libraries exist, they often require one to define the data to
visualize at a very low level (e.g. by providing the coordinates of the
triangles composing a surface mesh).  This is however very cumbersome and requires
lengthy custom code, while a teacher would strongly benefit from a simple
domain-specific widget requiring only minimal input. 
Widgets with this goal are provided
by OSSCAR as shown in the two examples in Fig.~\ref{fig:widgets}.  Panel (a)
shows a widget to plot the isosurfaces of molecular orbitals \cite{Grosso2013}.
It leverages the NGLview visualizer~\cite{Nguyen2017}, but exposes a simpler
interface to directly plot volumetric data associated to molecules.  Panel (b),
instead, shows a custom widget to compute and display interactively the first
Brillouin zone (BZ) \cite{Grosso2013} of a crystal.  It is based on the
JavaScript visualizer developed as part of the SeeK-path
library~\cite{seekpath}, and it exposes to teachers a very simple Python
interface: one just needs to provide the three real-space lattice vectors to
generate the BZ, where high-symmetry points are automatically displayed and
labelled, together with the suggested path to compute band structures.  Another
example of a custom widget developed in OSSCAR, not shown here, is an
interactive periodic table that allows users to select multiple chemical
elements (with each of them being in one of a range of possible states, e.g., to
select elements to be either included or excluded for searches and filtering).

We emphasize that developing a new widget might not be straightforward, as it
requires relatively advanced knowledge of both Python and JavaScript, as well as
experience with specific frameworks and libraries in the two languages.
However, once a \lib{Jupyter} widget has been developed and published, its use
is very straightforward, typically requiring only a couple of lines of Python
code.  Therefore, the custom OSSCAR widgets are a powerful catalyst that we
provide to facilitate and promote the creation of interactive notebooks with
powerful visualizations, and we expect to keep developing new ones with input and
contributions from the more expert user community.

\section{\label{sec:voila}Convert notebooks into web applications}

In the use cases A and B described in Sec.~\ref{sec:firstexample}, the primary
goal of the interactive visualizations is to deliver physics knowledge, with
less emphasis on programming and algorithms behind the notebook.  The code might
actually be distracting the first time a student interacts with the application,
and therefore we prefer to hide it in order to retain clarity in the
presentation.

A number of tools have been developed to convert \lib{Jupyter} notebooks into
shareable web applications, including appyters~\cite{Clarke2021},
nbinteract~\cite{Samuel2018}, bokeh~\cite{Bokeh}, \lib{Voila}~\cite{VoilaURL},
and appmode~\cite{AppModeURL}.  In OSSCAR we opted to use \lib{Voila}, a
subproject of Project Jupyter that can turn \lib{Jupyter} notebooks into live
standalone web applications, by executing the whole notebook and rendering only
the output cells into a web page format, while all of the source code is hidden.
For instance, Fig.~\ref{fig:2quantumwells} shows the page obtained for the
double quantum well after rendering the notebook with \lib{Voila}.
Additionally, \lib{Voila} keeps the Python code active (i.e., an active
connection is maintained between the web frontend and the so-called ``Python
kernel'' in \lib{Jupyter}).  This is crucial to allow the Python callbacks (see,
e.g., Fig.~\ref{fig:traitlets}) to be executed when the users interact with the
widgets.  Other solutions, instead, convert the notebook into a static webpage~\cite{nbconvert,JupyterBook}.
While this approach has the advantages of easier deployment (see also Sec.~\ref{sec:deployment}), it limits the interaction possibilities.
In addition, \lib{Voila}
supports the development of custom templates to modify the overall appearance of
web applications.  In OSSCAR, we have developed our own template that is also
shown in Fig.~\ref{fig:2quantumwells} (e.g., the header and footer with the
OSSCAR logo are part of this template) to provide a uniform and consistent look
and feel for all notebooks.

While the code can be completely hidden from the user and thus made fully
private using \lib{Voila}, we stress that in OSSCAR we strive to provide a
solution that fully complies with the Open Science principles: not only
regarding open availability of the applications for reuse in other classes, but
also releasing open source all code for inspection and reuse.  Therefore, all
source code of the OSSCAR notebooks is available as open source on GitHub, and
each notebook displays a link to it at the top of each page (see also Sec.~\ref{sec:docs}).

\section{\label{sec:deployment}Deployment on web/cloud servers}

The final aspect of making the notebooks available to a broad audience is their
deployment. This is crucial, as most users will not have the time (nor, often,
the expertise) to install locally Python, \lib{Jupyter} and all the dependencies
to run the notebooks.  Therefore, easy access to the applications with just a
web link becomes essential to make them straightforward to use.

However, deployment (especially if it has to be efficient) often comes with some
costs associated to it.  For instance, hosting content on most public cloud
services is not free, while self-hosted servers might also have a non-negligible
cost associated to the human time needed to maintain the service (perform system
security updates, recover after system crashes, \ldots).
Therefore, in OSSCAR we have investigated various solutions and, while we did
not find a single one that covers all requirements, we identify, use and suggest
three different (free) solutions to deploy and deliver the web applications,
depending on the student and teacher needs.  These are briefly described below,
highlighting pros and cons of each solution.

\subsection{\lib{mybinder.org}}
\lib{mybinder.org} is a website offering a free cloud solution to deploy
\lib{Jupyter} notebooks, and has been already used to serve course content to students~\cite{Kim2021}.
\lib{mybinder.org} allows one to generate a unique URL
associated to a GitHub repository that contains notebooks and code.  When a user
opens the link, \lib{mybinder.org} automatically fetches the code and runs it in
an isolated environment for each user (using \lib{Docker}~\cite{DockerURL}
containers behind the scene) so that each user does not interact with others
using the application at the same time.

\begin{figure}[tb]
   \centering\includegraphics[width=7.5cm]{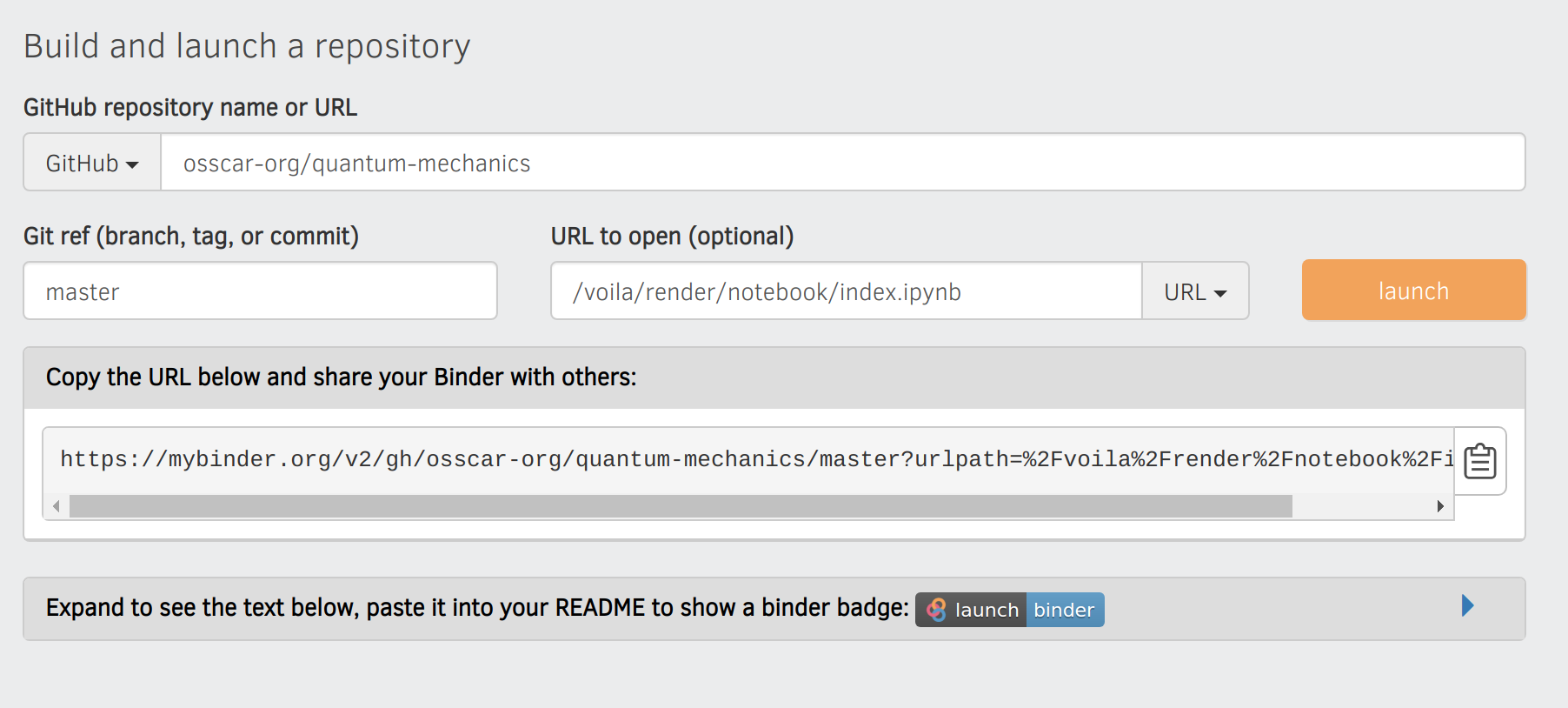}
    \caption{\label{fig:binder} Input GitHub details to generate the \lib{mybinder.org} link.} 
\end{figure}

\lib{mybinder.org} requires minimal effort for teachers.  One first has to
create a public GitHub repository with the notebooks and some basic
configuration files that, as detailed in their documentation, primarily consist
of a list of Python dependencies that need to be installed in order to make the
notebooks functional.  Then, on the \lib{mybinder.org} homepage, one can easily
obtain a unique link (that can be distributed to students, or published on a
webpage) to access the deployed application.  The link can be obtained simply by
providing the GitHub repository name, the Git branch name and the notebook URL
(see Fig.~\ref{fig:binder}).  Furthermore, using \lib{Voila} together with
\lib{mybinder.org} is very straightforward: one can just declare
\lib{Voila} among the dependencies, and then prepend the string
\texttt{/voila/render/} to the notebook URL to trigger the \lib{Voila} extension
at load time.

Being free, open and requiring almost zero maintenance effort, this service is extremely useful, but there are two shortcomings.  The most critical one is
that every time that the page is loaded, the initialization might take a
significant amount of time (from a few tens of seconds to some minutes).  This
might be problematic for the needs of the two use cases described in
Sec.~\ref{sec:firstexample}, and in particular for use case B, where a teacher
might want to use the application only for a very short amount of time.  In
addition, being a free service, the computing power is also very limited, which
can be an issue for sophisticated notebooks performing advanced simulations.
Nevertheless, we strongly encourage any teacher developing a notebook to provide
a \lib{mybinder.org} link in their homepage, as this makes the notebooks
immediately accessible (even if with a lag of a few tens of seconds) to any web
user, without any setup needed.

\subsection{\lib{dokku} deployment}
In order to speed up the startup time of each notebook, we also deploy the
OSSCAR notebooks on custom resources using an open-source software called
\lib{dokku}~\cite{Dokku}.  \lib{dokku} is an extensible Platform-as-a-Service
software that makes deployment of applications extremely easy.  In particular,
one just needs to place all their code and notebooks inside a Git repository,
and push the content to the \lib{dokku} server to update the deployed version.

Similarly to \lib{mybinder.org}, \lib{dokku} transparently creates a
\lib{Docker} container. This container is, however, the same one for all users
and user isolation is obtained thanks to \lib{Voila}. This requires special care
when implementing the notebooks to avoid unexpected interactions between users,
e.g., if files with the same name are generated on the server.  The startup time,
however, is considerably reduced, typically to 5 seconds or less.

Unfortunately, this solution also has some shortcomings. In addition to having
to understand the deployment model to prevent unexpected interaction among
different users, installing and maintaining a \lib{dokku} server requires the
availability of an online server (that might not be free) and most importantly
it requires expertise in managing and deploying web servers.  In our case, we
leverage the \lib{dokku} service provided by the Materials Cloud
portal~\cite{Talirz2020}, with servers hosted at the Swiss National
Supercomputing Center (CSCS), that kindly provides the resources to host the
OSSCAR applications.  For instance, the applications for quantum mechanics
described later in Sec.~\ref{sec:example-apps} can be accessed at the address
\url{https://osscar-quantum-mechanics.materialscloud.io}.  If, however, a
teacher does not have access to such a deployment, this solution might not be
viable (we note, however, that similar hosted commercial solutions exist, such
as \lib{heroku.com} for instance, that might have a free tier for small
non-commercial applications).

\subsection{Institutional \lib{JupyterHub} servers}
Because of the widespread adoption of \lib{Jupyter}, many universities, research
centers and computer centres are now offering to their users (teachers,
students, researchers) access to locally hosted \lib{JupyterHub} installations.
\lib{JupyterHub} is an open server facilitating the provision of multi-user
access to notebooks.

This solution could be ideal for courses given at universities where this
service exists and all students have access to it.  The added benefit of this
deployment approach is that each student has access to their persistent home
folder, where they can not only install and use the applications, but also
easily modify the code and run the modified versions, possibly contributing back
their changes to the original repository.  This solution is therefore
particularly suitable if the teachers want to encourage the students to modify
and adapt the code of the notebooks.

\begin{figure}[tb]
   \centering\includegraphics[width=7.5cm]{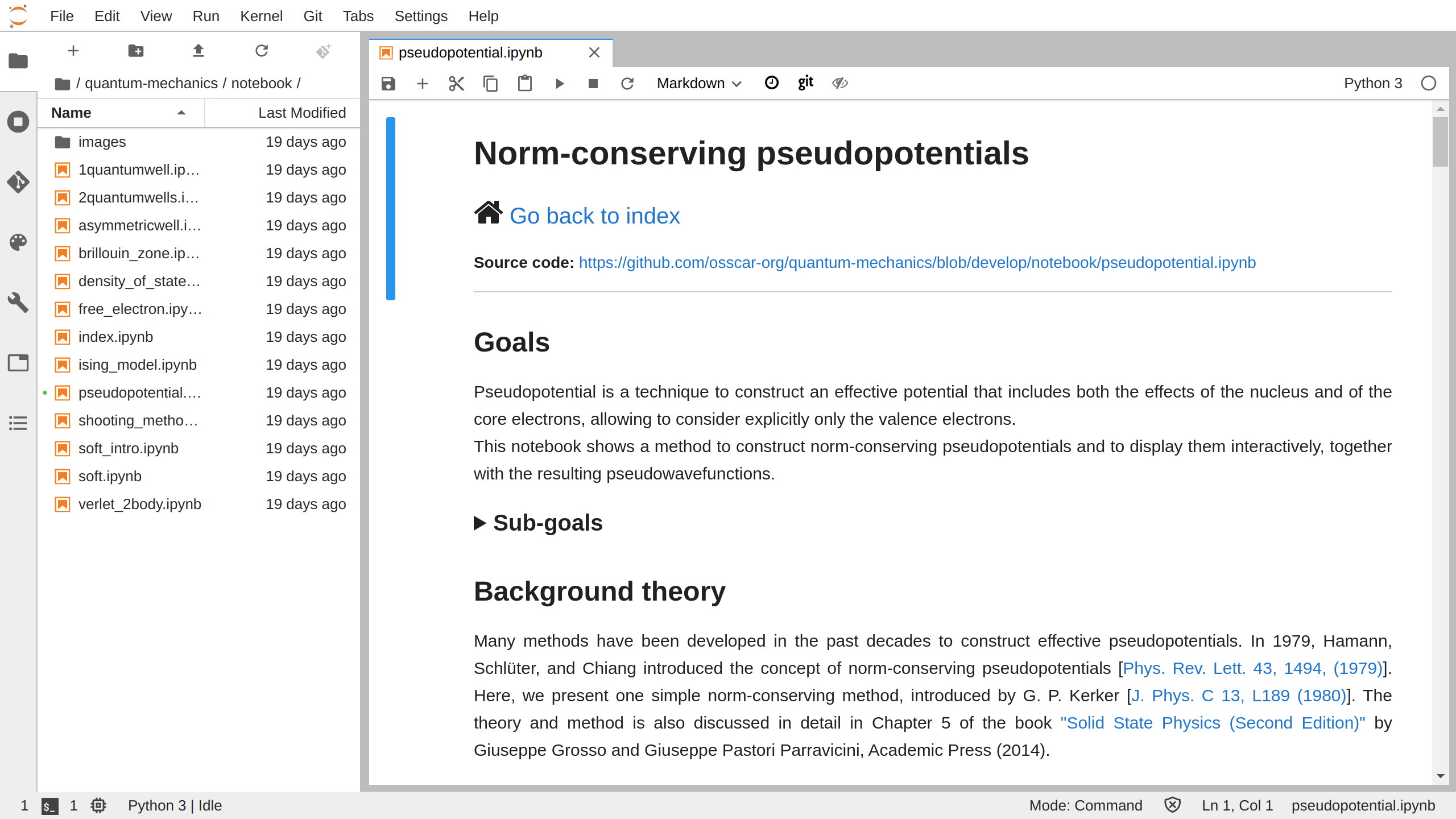}
    \caption{\label{fig:noto} An example of how an OSSCAR notebook is displayed in the \lib{JupyterLab} interface, in this specific example available via the EPFL NOTO platform.} 
\end{figure}

One example of such an institutional \lib{JupyterHub} is the NOTO platform at
EPFL (\url{https://noto.epfl.ch}).  We show how a notebook appears inside the
\lib{JupyterLab} interface provided by NOTO in Fig.~\ref{fig:noto}, but we
stress that many more universities are already providing a similar service, also
thanks to the fact that \lib{JupyterHub} and \lib{JupyterLab} are open source
and officially supported as part of Project Jupyter.

We finally mention that in terms of use cases covered, also Google Colab~\cite{googlecolab} can be considered to fall within this category, where rather than institutional credentials one would need instead a Google login.

\section{\label{sec:example-apps}OSSCAR notebooks for computational science}

\begin{figure*}[tb]
   \centering\includegraphics[width=1.0\linewidth]{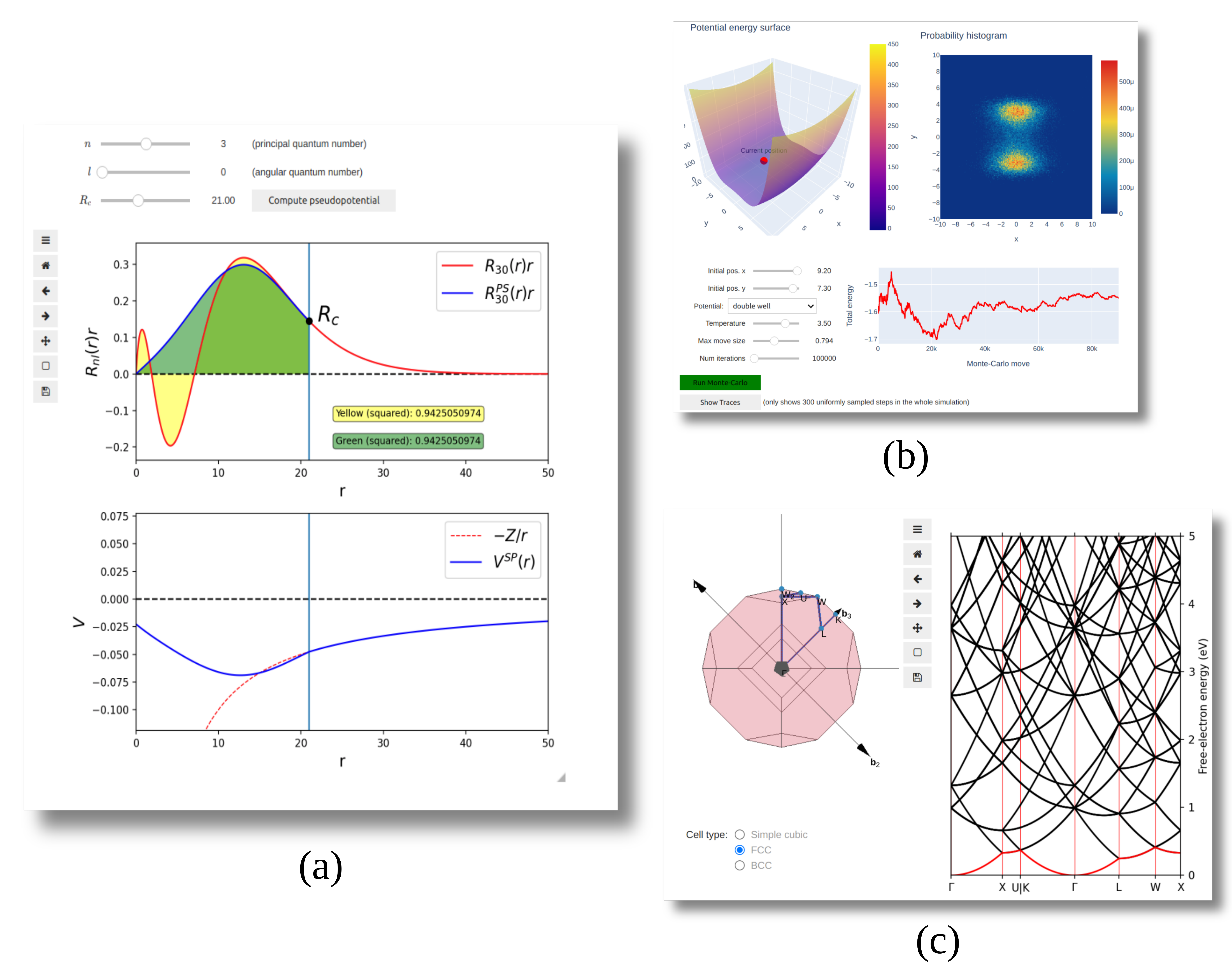}
    \caption{\label{fig:notebooks} Three interactive web applications developed
    in OSSCAR and used to complement teaching in classes of computational simulations.
    The web applications are hosted at 
    \url{https://osscar-quantum-mechanics.materialscloud.io}.
    a) Construction of
    a norm-conserving pseudopotential.
    b) Monte-Carlo simulations to sample the canonical distribution of a given potential energy surface.
    c) Electronic band structure and first BZ of an empty-lattice free-electron crystal.} 
\end{figure*}

In addition to proposing guidelines for best practices in developing open
teaching content, and developing custom widgets for computational-science
content, one of the main goals in OSSCAR is to generate open and free learning
content in the broad domain of computational physics, chemistry, and materials
science.

OSSCAR currently offers a number of interactive notebooks covering the topics of
quantum mechanics, band theory of crystals, statistical mechanics, and molecular
dynamics.  The choice of the topics stems primarily from the content of two
courses taught by some of the authors at EPFL (``Computational methods in
molecular quantum mechanics'' and ``Atomistic and quantum simulations of
materials'').  The notebooks have been already used in the past year with very
positive feedback from students.  In particular, anonymous surveys were
conducted at the end of one of the courses.  The results indicate that for the
majority of students (over 70\%) the inclusion of interactive visualizations
during the classes was both motivating, and helped them to better understand the
core course concepts by being actively engaged in the learning process.
Furthermore, the same proportion of students also accessed and used the
interactive visualizations to improve their understanding after the lectures,
while revising the course content (use case A in Sec.~\ref{sec:firstexample}).
Students also provided valuable feedback for improvement of the
notebooks, and, in both classes, some even demonstrated interest in generating
new educational content using the same OSSCAR approach.

Without aiming at presenting an exhaustive list, in the following we briefly
show some selected examples of applications developed in OSSCAR, to demonstrate
with practical examples the general concepts and technologies (such as the
widgets) described earlier. More applications are available online, and we
expect the list to continue growing in the future.  The source codes of all
notebooks are available on the GitHub repository at
\url{https://github.com/osscar-org/quantum-mechanics} and can be directly
inspected on our \lib{dokku} server at
\url{https://osscar-quantum-mechanics.materialscloud.io}.

Fig.~\ref{fig:notebooks} presents three different OSSCAR notebooks.  As
mentioned in \ref{app:2D3D-plotting}, we use \lib{matplotlib} to render 2D
interactive figures.  This is the case, for instance, for the two plots in
Fig.~\ref{fig:notebooks}(a), an application illustrating the construction of
norm-conserving pseudopotentials~\cite{Grosso2013}.  The two panels display the
hydrogen-atom potential and the pseudopotential that was generated (bottom
panel) and one of the wavefunctions and the corresponding pseudo-wavefunction
(top panel).  Users can select the principal quantum number $n$ and the angular
quantum number $l$ of the wavefunction in the controllers region, as well as the
cutoff radius $R_c$ determining the core region.

In Fig.~\ref{fig:notebooks}(b) we instead show an application that illustrates
the use of Monte-Carlo simulations with the Metropolis--Hastings algorithm
\cite{frenkel2001understanding, teukolsky1992numerical} to sample the canonical
distribution at a given temperature $T$ for a potential that can be selected among various possibilities (a 2D double-well potential is selected and shown in
the figure).  Students can set the starting coordinates $(x, y)$ of the
simulation, select the temperature $T$, and tune the simulation parameters
(maximum move size and total number of iterations), to investigate both physical
(temperature and potential-barrier height) and numerical effects on the
efficiency and ergodicity of the simulation. This application displays various
types of figures: a 3D visualization of the potential energy surface (top left,
displayed with \lib{plotly} as discussed in \ref{app:2D3D-plotting}), the
probability histogram as a color map obtained from the simulation (top right,
displayed using \lib{matplotlib}), and the total energy as a function of the
Monte-Carlo move (bottom right).  This notebook is also an example of the
Multiple Representation Principle, where complementary representations of
related quantities are displayed to facilitate learning, thanks to the different
informational content of each of them~\cite{Ainsworth2014}. 

Finally, Fig.~\ref{fig:notebooks}(c) shows an application to compute and show
the band structure and BZ of a simple empty-lattice
free-electron crystal (for simple, face-centered and body-centered cubic
lattices)~\cite{Grosso2013}.  The notebook can be used to explain the
concept of reciprocal space, discuss how band structure paths are selected, and
compare band structures of actual materials with the free-electron case.  While
the band structure plot (on the right) uses the same \lib{matplotlib} library
that we employ for 2D plots, the left part uses the custom OSSCAR BZ visualizer
already described in Fig.~\ref{fig:widgets}(b).

\begin{figure*}[tb]
   \centering \includegraphics[width=0.7\linewidth]{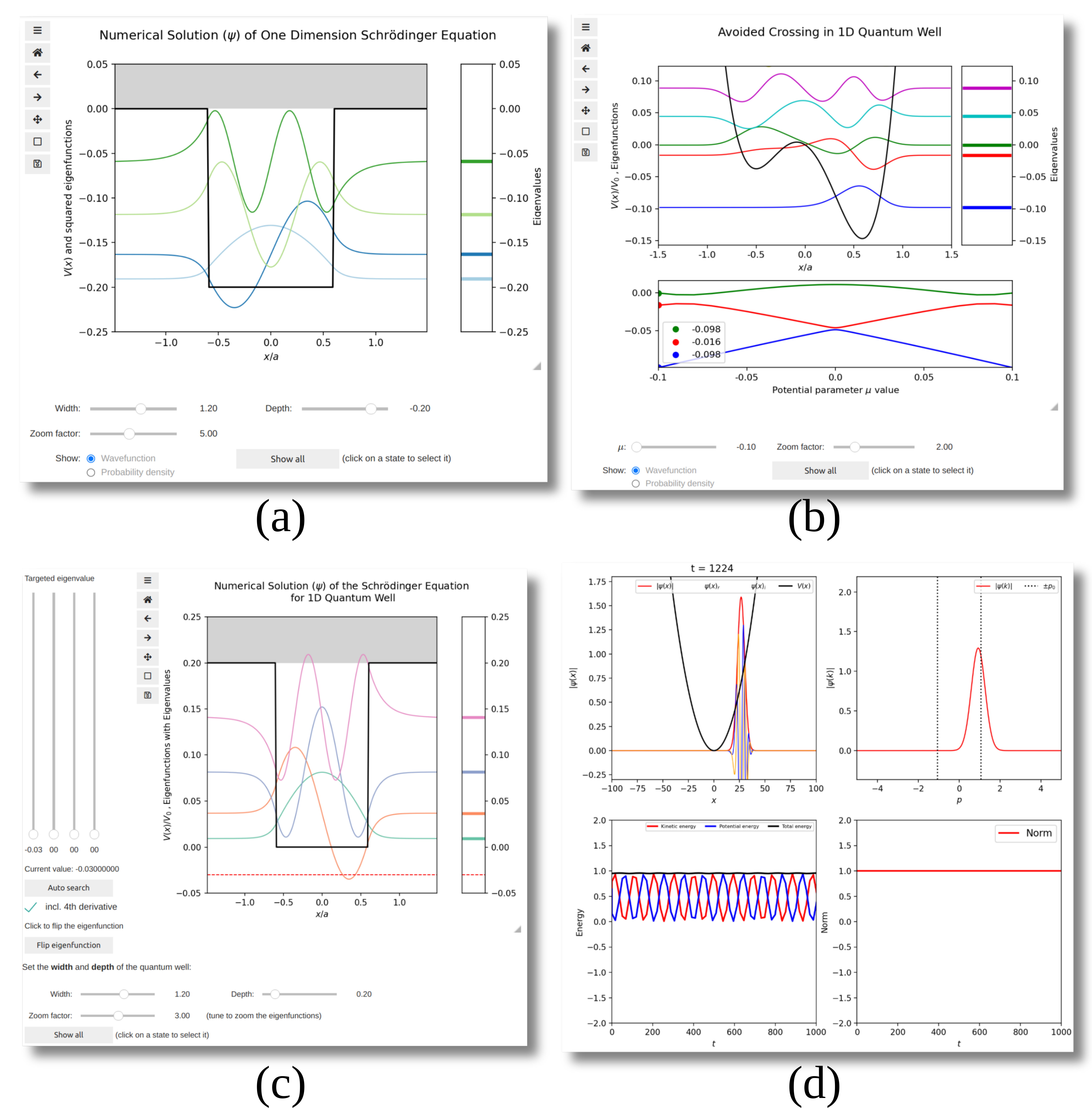}
    \caption{\label{fig:quantum} Four OSSCAR applications of different levels (from basic to advanced) addressing the general topic of the solution of the Schrödinger equation, to demonstrate how an advanced topic can be split into smaller self-contained notebooks.
    a) Numerical solution of the 1D time-independent Schrödinger equation for a single quantum-well potential.
    b) Avoided crossing in a 1D asymmetric quantum well.
    c) Explanation of the shooting method and Numerov algorithm to solve the Schrödinger equation.
    d) Numerical solution of the 1D time-dependent Schrödinger equation using the split-operator Fourier transform (SOFT) method and time evolution of a wavepacket.} 
\end{figure*}

\subsection{A library of focused self-contained applications}

From a learning perspective, we strive to design each web application to be
self-contained and focused on conveying one single core concept.  When more
complex concepts need to be explained, we prefer and suggest splitting the
content into a sequence of propaedeutic smaller notebooks, each focusing on a
single (sub)topic.  This makes each application easy to use even without teacher
supervision, and the series of notebooks guides the students in a progressive
learning process.  To demonstrate this modular approach, we show in
Fig.~\ref{fig:quantum} four different examples of notebooks focusing on basic
quantum-mechanical concepts, in addition to the double quantum well already
presented in Fig.~\ref{fig:2quantumwells}.

The first notebook of the series is shown in Fig.~\ref{fig:quantum}(a), focusing
on the numerical solution of the Schrödinger equation for a single 1D finite
square-well potential \cite{izaac2018computational}.  Being one of the simplest
quantum models, it allows students to start familiarizing themselves with the
visualization of quantum eigenstates, and to inspect the effect of quantum
confinement \cite{shankar2012principles}.

The second notebook in the series is the double quantum well already discussed
in Fig.~\ref{fig:2quantumwells}: having two wells, it allows students to
investigate their interplay and the effect of quantum tunneling.  A slightly
more advanced model is presented in the notebook of Fig.~\ref{fig:quantum}(b),
where a 1D asymmetric quantum-well system is now proposed, described by the
expression $V(x) = x^4 - 0.6x^2 + \mu x$.  The parameter $\mu$ can be tuned via
a slider to determine the amount of asymmetry between the two wells.  The lower
panel, showing the eigenenergies of the three lowest states in the system as a
function of $\mu$, helps students to focus on the phenomenon of avoided crossing
\cite{tannor2007introduction}.

Fig.~\ref{fig:quantum}(c) goes back to the same single quantum-well model of
Fig.~\ref{fig:quantum}(c).  However, the teaching focus in this case is not on
the solutions of the equation, but on the algorithm to obtain them.  In
particular, this notebook aims at describing the shooting method using Numerov's
algorithm \cite{thijssen2007computational}.  The vertical sliders on the left
allow one to choose a ``guess'' energy, that will be used to determine a
wavefunction with the correct boundary condition for $x\to-\infty$ (vanishing
wavefunction).  However, only the actual eigenvalues of the system will return a
wavefunction that fulfills the vanishing boundary condition also at
$x\to+\infty$.  The students can then try various values to understand how the
algorithm works.  For convenience, we also provide an ``Auto search'' button,
which implements the full algorithm and aids in quickly finding the correct
solutions.

Finally, the most advanced notebook is shown in Fig.~\ref{fig:quantum}(d).
Unlike the previous applications (solving the time-independent Schrödinger
equation), this notebook demonstrates the solution of the time-dependent
Schrödinger equation using the split-operator Fourier transform (SOFT) numerical
method~\cite{tannor2007introduction,fleck1976time}.  After choosing a potential
energy shape, a wavepacket is constructed and its time evolution is computed and
displayed.  The various panels enable the monitoring of the wavepacket evolution
in real and reciprocal space (top panels), as well as the kinetic and potential
energy of the packet and the conservation of the total energy (bottom left
panel) and of the norm of the wavefunction (bottom right panel) to inspect the
robustness of the algorithm.

\section{\label{sec:docs}Documentation, tutorials, and source code to engage teachers and students}

As we mentioned earlier, the overarching goal of OSSCAR is to encourage an
open-science approach for education, encouraging other teachers to develop their
own educational interactive web applications and inviting them to share them on
this platform.

The previous sections describe a number of strategies that are all instrumental
to this objective.  As an additional effort toward this core goal, in OSSCAR we
also provide online extensive documentation and tutorials, accessible from the
OSSCAR homepage (\url{https://www.osscar.org}).  The documentation, in
conjunction with the library of notebooks, assists and encourages teachers in
developing further teaching content.  Indeed, the existing notebooks are all
released with open-source licenses and hosted on GitHub repositories of the
OSSCAR organization (\url{https://github.com/osscar-org}). These serve as
examples for development of new applications. In addition, the repositories not
only contain the notebooks with the source code of all applications, but also
all configuration files needed for deployment on, e.g., \lib{mybinder.org} or on
\lib{dokku}. Therefore, each repository is a complete template to develop a new
web application. Teachers can extract the notebooks or just the configuration
files, modify them for their individual needs, and, if they wish, contribute new notebooks for
different courses.

Furthermore, as discussed in Sec.~\ref{sec:firstexample}, we provide a direct
link at the top of each application to directly access the source code, aiming
at multiple objectives.  First, interested students (after having interacted
with the notebook) can inspect the code to see which algorithms have been used
to solve the equations, and possibly adapt the codes and algorithms to gain an
even deeper understanding of the subject.  Second, code access encourages both
students and other teachers to provide feedback and improvements via GitHub
issues and pull requests, in a fully collaborative and open approach and in the
spirit of Open Science.  Third, by engaging the students in the preparation of
the content, teachers can implement and encourage in their courses approaches of
peer instruction and cooperative learning, that have been shown to increase
student engagement and understanding~\cite{Crouch2001}.

\section{Conclusions}

We presented OSSCAR, an open web-based platform for educational content. OSSCAR
provides a collaborative environment where teachers can easily develop, deploy,
and distribute to students interactive notebooks that facilitate scientific
learning via visualization, examples, and numerical experimentation.  The
platform aims at hosting a growing number of modules, each tackling a specific
topic and with the potential to be combined and organized in multiple ways,
based on the needs of each class.  This free online library will hopefully
provide a set of ``off-the-shelf'' tools to complement classical teaching, and
attract contributions by a large community of teachers recognizing the advantage
of sharing and improving over duplicating.  New content is welcome and can be
easily created in the OSSCAR environment, that relies on user-friendly and common
languages and software, such as Python and \lib{Jupyter}, as the key development
tools.  Easy deployment of the notebooks is achieved by their automatic
conversion into web applications via the \lib{Voila} software, and then by
hosting them on existing or custom web cloud solutions.  Students can thus
access the material directly via their web browser, avoiding the need of
tailored installations for each individual course.  They learn by performing
specific tasks, solving exercises, and -- importantly -- experimenting in real
time with the interactive content of the notebooks.  Further information on the
OSSCAR project and the documentation can be found on the project web page:
\url{https://www.osscar.org}.  Examples, custom widgets and templates for the
development of OSSCAR web applications are available on GitHub at
\url{https://github.com/osscar-org}.

\section*{Acknowledgements}

We acknowledge financial support from the EPFL Open Science Fund via the OSSCAR
project.  We acknowledge CECAM for dedicated OSSCAR dissemination activities.
We acknowledge the NCCR MARVEL (a National Centre of Competence in Research, funded by the Swiss National Science Foundation, grant No. 182892), the
European Centre of Excellence MaX ``Materials design at the Exascale'' (grant
No. 824143) and the European Union's Horizon 2020 research and innovation
programme under grant agreement No. 957189 (BIG-MAP), also part of the BATTERY
2030+ initiative under grant agreement No. 957213, for their support in the
deployment of the applications on the Materials Cloud (via \lib{dokku}).  The
authors are grateful to Michele Ceriotti, Nicola Marzari, Ignacio Pagonabarraga
and Berend Smit for fruitful discussions, C\'ecile Hardebolle for feedback and
useful discussions on how to better design the notebooks to increase their
learning effectiveness, Pierre-Olivier Vall\`es for the support for the
deployment on the EPFL NOTO \lib{JupyterHub} platform, Guoyuan Liu for
implementing two notebooks, and the students of the courses where the OSSCAR
content was used for providing valuable feedback on style and content.

\appendix

\section{\label{app:speedup}Strategies to speed up Python simulations}

The first and foremost approach to accelerate simulations is to
optimize, rethink or adapt the algorithm.  However, the use of packages such as
\lib{NumPy}~\cite{Harris2020} and \lib{SciPy}~\cite{Virtanen2020} (nowadays
standard dependencies of a vast majority of scientific Python code) helps in
making a wide range of complex but common operations (such as matrix operations,
advanced optimization routines, ...) easy to use and as efficient as compiled
languages, since internally the core computational routines are implemented in
C, C++ or Fortran.  Furthermore, other technologies and libraries exist to speed
up Python code.  We mention here only few examples, used in some OSSCAR
notebooks: the \lib{Numba}~\cite{Lam2015} package, to write Python codes using
only simple types and arrays and to convert them to C codes on the fly with a
just-in-time (JIT) compiler; and the \lib{Cython}~\cite{Behnel2011} package (and
\lib{f2py}~\cite{f2py}, now part of \lib{NumPy}) to write computationally
expensive routines directly in C (and Fortran, respectively), and then call
those from Python.

\section{\label{app:2D3D-plotting}Libraries for visualization and plotting}
One of the most typical tools for visualization in scientific applications are
plots in two or three dimensions. Many libraries for such common plots are
available in Python and are interfaced with \lib{Jupyter}, including
\lib{matplotlib}~\cite{Hunter2007}, \lib{plotly}~\cite{Jon2018},
\lib{bqplot}~\cite{bqplotURL} or \lib{ParaView}~\cite{Ahrens2005}.

With the approach that we describe in this paper, we do not limit or prescribe
which libraries can be employed to develop new applications.  Nevertheless, we
made some considered decisions on our first choice libraries, trying to select
the smallest set of different libraries that can cover use cases most commonly
encountered, are fast enough for large datasets, and have wide community
support. By favoring reuse of the same libraries in multiple notebooks, we
provide a consistent user experience to students, and at the same time the
notebooks become a suite of examples of how to interact with the chosen
libraries.

In particular, in the OSSCAR notebooks we use \lib{matplotlib} as the main
plotting package for two-dimensional (static, animated and interactive) plots,
such as charts or color plots. As an example, the interactive figures in
Fig.~\ref{fig:2quantumwells}(c) are produced using \lib{matplotlib}.  For the
purpose of illustration, in the following we show a short, but fully
functioning, Python code to demonstrate the simple syntax required to generate
basic plots~\footnote{Naturally, slightly longer code is needed to achieve more
refined results, e.g. to change the panels aspect ratio, the color of the plots,
etc.}.

We can generate two panels side by side with the following code:
\begin{lstlisting}[language=Python]
%matplotlib notebook
import pylab as plt
fig, axes = plt.subplots(1, 2)
\end{lstlisting}
where the first line enables interactive plots, the second imports the main
plotting module of the \lib{matplotlib} library, and the third generates the
empty panels.

The code defines \texttt{axes} as a list of two subplots, with \texttt{axes[0]}
being the left one and \texttt{axes[1]} the right one.  We can now plot  a curve
on the left panel with:
\begin{lstlisting}[language=Python]
axes[0].plot(x, y)
\end{lstlisting}
In this instruction, \texttt{x} is a Python list of $x$ coordinates of each of
the points, and \texttt{y} the corresponding list of $y$ coordinates.
Similarly, we can use \texttt{axes[1].plot} to plot curves on the right panel.

We also mention that \lib{matplotlib} figures support dynamical updates.  For
instance, one can remove all curves from the left panel (e.g., when redrawing
the figure if a controller value is changed by the user) with:
\begin{lstlisting}[language=Python]
axes[0].clear()
\end{lstlisting}
or replace the $y$ data of the first curve (\texttt{lines[0]}) of the right
panel (\texttt{axes[1]}) with the data in the list \texttt{new\_y} via:
\begin{lstlisting}[language=Python]
axes[1].lines[0].set_ydata(new_y)
\end{lstlisting}
    
In addition, the \lib{matplotlib} library offers a large number of different
types of plots, the possibility of showing text and annotations in the plots and
more generally to customize almost any aspect of the plots. It is  also possible
to interact with the plots and detect, for instance, the position of a mouse
click. 

While \lib{matplotlib} can also generate three-dimensional plots, in our
experience its performance was often not good enough for smooth and pleasant
interaction (e.g., noticeable latency when rotating or zooming).  Therefore, for
three-dimensional plots we use instead the \lib{plotly} library, that showed
better performance. An example is given by the plot of the potential energy
surface in Fig.~\ref{fig:notebooks}(b).

\section{\label{app:widgets-example}An example widget and instantaneous reaction to events}
\begin{figure}[tbp]
   \centering \includegraphics[width=7.5cm]{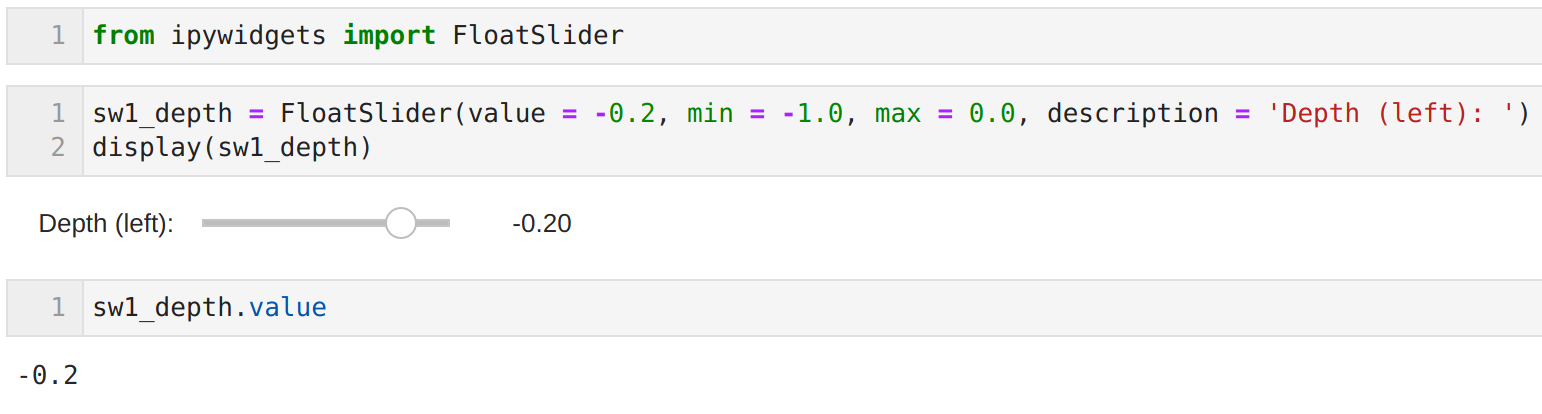}
    \caption{\label{fig:slider}Creation of a slider widget (from the
        \lib{ipywidgets} package) to enable the user to control a (floating
        point) numerical value.  The parameters passed to the initializer allow
        to decide the default initial \texttt{value} and the range
        (\texttt{min}, \texttt{max}) of allowed inputs, together with a textual
        \texttt{description} of the widget.  In subsequent cells, it is possible
        (from Python) to retrieve the current value of the widget, or set its
        value programmatically, using the \texttt{.value} property.
    } 
\end{figure}

Each widget can be created via Python code directly in a
notebook cell.  For example, Fig.~\ref{fig:slider} shows the code used to create
the slider to control the depth of the first square well in
Fig.~\ref{fig:2quantumwells}, and retrieve its value programmatically from
Python.

Beside being able to check the value of the slider in specific points of the
code, an essential part of the interactivity comes from a very small time delay
between user actions (button clicks, change of the value of a slider, \ldots)
and the adaptive reaction of the notebook.  In OSSCAR, this is achieved using
the \lib{traitlets} library~\cite{Traitlets}.  In particular, every time the
attributes of a widget are modified, the widget emits an event of type
``change''.  We can then define callback functions that are triggered every time
there is a change, and bind them to the event using the \texttt{observe} method
of the widget.  For instance, in Fig.~\ref{fig:traitlets} we show a code snippet
defining a callback function \texttt{slider\_value\_change} to replot the
function in the figure generated using \texttt{axes[0].plot} discussed in \ref{app:2D3D-plotting} after changes triggered by the \texttt{sw1\_depth} slider.

\begin{figure}[tb]
\begin{lstlisting}[language=Python]
# define the callback function to replot
def slider_value_change(c):
    ...
    axes[0].plot(x, V)
    ...

# bind the callback, so that it is triggered every time
# that the value of sw1_depth changes
sw1_depth.observe(slider_value_change, names="value")
\end{lstlisting}
\caption{\label{fig:traitlets}A minimal example of how to define a callback and bind it to any change of value of the slider of Fig.~\ref{fig:slider}.}
\end{figure}

Finally, we mention that events can be associated to any widget, including plots,
thus allowing to tune the value of certain parameters not only from the controllers
section, but also by clicking directly on the visualizations (and, in this case,
adapting the value of the controllers accordingly).  This approach allows for
the implementation of reciprocative dynamic linking between components, that
has been shown to improve representational competence in
students~\cite{Patwardhan2017}.


\begin{thebibliography}{10}
    \expandafter\ifx\csname url\endcsname\relax
      \def\url#1{\texttt{#1}}\fi
    \expandafter\ifx\csname urlprefix\endcsname\relax\def\urlprefix{URL }\fi
    \expandafter\ifx\csname href\endcsname\relax
      \def\href#1#2{#2} \def\path#1{#1}\fi
    
    \bibitem{deJong2013}
    T.~de~Jong, M.~C. Linn, Z.~C. Zacharia, Physical and virtual laboratories in
      science and engineering education, Science 340~(6130) (2013) 305--308.
    
    \bibitem{QuantumPhysicsOnline}
    M.~Joffre, Quantum physics online,
      \url{https://www.quantum-physics.polytechnique.fr} (2019).
    
    \bibitem{softmatterdemos}
    F.~Smallenburg, L.~Filion, R.~M. Alkemade, A.~Ulug\"ol, Soft matter demos,
      \url{https://www.softmatterdemos.org} (2022).
    
    \bibitem{nanohub}
    {Network for Computational Nanotechnology}, Nanohub, \url{https://nanohub.org}
      (2022).
    
    \bibitem{LabXchange}
    {The LabXchange team}, Labxchange, \url{https://www.labxchange.org} (2022).
    
    \bibitem{Vallejo2022}
    W.~Vallejo, C.~D{\'{\i}}az-Uribe, C.~Fajardo, Google colab and virtual
      simulations: Practical e-learning tools to support the teaching of
      thermodynamics and to introduce coding to students, {ACS} Omega (Feb. 2022).
    
    \bibitem{ijerph18052672}
    H.~Kawasaki, S.~Yamasaki, Y.~Masuoka, M.~Iwasa, S.~Fukita, R.~Matsuyama, Remote
      teaching due to covid-19: An exploration of its effectiveness and issues,
      International Journal of Environmental Research and Public Health 18~(5)
      (2021).
    
    \bibitem{youmans2020}
    M.~Youmans, Going remote: How teaching during a crisis is unique to other
      distance learning experiences, J. Chem. Educ. 97~(9) (2020) 3374–--3380.
    
    \bibitem{Zalat2021}
    M.~M. Zalat, M.~S. Hamed, S.~A. Bolbol, The experiences, challenges, and
      acceptance of e-learning as a tool for teaching during the covid-19 pandemic
      among university medical staff, PLOS ONE 16~(3) (03 2021).
    
    \bibitem{10.3389/feduc.2021.638470}
    Z.~Almahasees, K.~Mohsen, M.~O. Amin, Faculty’s and students’ perceptions
      of online learning during covid-19, Frontiers in Education 6 (2021).
    
    \bibitem{PhysRevPhysEducRes.17.020111}
    J.~R. Hoehn, M.~F.~J. Fox, A.~Werth, V.~Borish, H.~J. Lewandowski, Remote
      advanced lab course: A case study analysis of open-ended projects, Phys. Rev.
      Phys. Educ. Res. 17 (2021) 020111.
    
    \bibitem{doi:10.1021/acs.jchemed.1c00655}
    R.~Kobayashi, T.~P.~M. Goumans, N.~O. Carstensen, T.~M. Soini, N.~Marzari,
      I.~Timrov, S.~Poncé, E.~B. Linscott, C.~J. Sewell, G.~Pizzi, F.~Ramirez,
      M.~Bercx, S.~P. Huber, C.~S. Adorf, L.~Talirz, Virtual computational
      chemistry teaching laboratories—hands-on at a distance, Journal of Chemical
      Education 98~(10) (2021) 3163--3171.
    
    \bibitem{Yakutovich2021}
    A.~V. Yakutovich, K.~Eimre, O.~Schütt, L.~Talirz, C.~S. Adorf, C.~W. Andersen,
      E.~Ditler, D.~Du, D.~Passerone, B.~Smit, N.~Marzari, G.~Pizzi, C.~A.
      Pignedoli, Aiidalab – an ecosystem for developing, executing, and sharing
      scientific workflows, Computational Materials Science 188 (2021) 110165.
    
    \bibitem{jupyter}
    T.~Kluyver, B.~Ragan-Kelley, F.~P{\'e}rez, B.~Granger, M.~Bussonnier,
      J.~Frederic, K.~Kelley, J.~Hamrick, J.~Grout, S.~Corlay, P.~Ivanov, D.~Avila,
      S.~Abdalla, C.~Willing, J.~development team, Jupyter notebooks - a publishing
      format for reproducible computational workflows, in: F.~Loizides, B.~Scmidt
      (Eds.), Positioning and Power in Academic Publishing: Players, Agents and
      Agendas, IOS Press, Netherlands, 2016, pp. 87--90.
    
    \bibitem{JupyterURL}
    {Project Jupyter}, Jupyter, \url{https://docs.jupyter.org} (2022).
    
    \bibitem{Granger2021}
    B.~E. Granger, F.~Pérez, Jupyter: Thinking and storytelling with code and
      data, Computing in Science Engineering 23~(2) (2021) 7--14.
    
    \bibitem{JupyterLabURL}
    {Project Jupyter}, Jupyterlab, \url{https://github.com/jupyterlab/jupyterlab}
      (2022).
    
    \bibitem{VoilaURL}
    {Voila Development Team}, Voila,
      \url{https://github.com/voila-dashboards/voila} (2022).
    
    \bibitem{Grosso2013}
    G.~Grosso, G.~Pastori~Parravicini, {Solid State Physics}, Academic Press,
      Oxford, UK, 2013.
    
    \bibitem{Hake1998}
    R.~R. Hake, Interactive-engagement versus traditional methods: A
      six-thousand-student survey of mechanics test data for introductory physics
      courses, American Journal of Physics 66~(1) (1998) 64--74.
    
    \bibitem{Crouch2004}
    C.~Crouch, A.~P. Fagen, J.~P. Callan, E.~Mazur, Classroom demonstrations:
      Learning tools or entertainment?, American Journal of Physics 72~(6) (2004)
      835--838.
    
    \bibitem{Freeman2014}
    S.~Freeman, S.~L. Eddy, M.~McDonough, M.~K. Smith, N.~Okoroafor, H.~Jordt,
      M.~P. Wenderoth, Active learning increases student performance in science,
      engineering, and mathematics, Proceedings of the National Academy of Sciences
      111~(23) (2014) 8410--8415.
    
    \bibitem{Perkel2015}
    J.~M. Perkel, Programming: Pick up python, Nature 518~(7537) (2015) 125--126.
    
    \bibitem{Pip}
    {PyPi Development Team}, Pip, \url{https://pypi.org/project/pip} (2021).
    
    \bibitem{Conda}
    {Conda Development Team}, Conda, \url{https://docs.conda.io} (2021).
    
    \bibitem{Perkel2018}
    J.~M. Perkel, By jupyter, it all makes sense, Nature 563~(7729) (2018)
      145--146.
    
    \bibitem{Barba2019}
    L.~A. Barba, L.~J. Barker, D.~S. Blank, J.~Brown, A.~B. Downey, T.~George,
      L.~J. Heagy, K.~T. Mandli, J.~K. Moore, D.~Lippert, K.~E. Niemeyer, R.~R.
      Watkins, R.~H. West, E.~Wickes, C.~Willing, , M.~Zingale, Teaching and
      learning with jupyter, \url{https://jupyter4edu.github.io/jupyter-edu-book}
      (2019).
    
    \bibitem{Weiss2017}
    C.~J. Weiss, Scientific computing for chemists: An undergraduate course in
      simulations, data processing, and visualization, Journal of Chemical
      Education 94~(5) (2017) 592--597.
    \newblock \href {https://doi.org/10.1021/acs.jchemed.7b00078}
      {\path{doi:10.1021/acs.jchemed.7b00078}}.
    
    \bibitem{Gilbert2008}
    J.~K. Gilbert, M.~Reiner, M.~Nakhleh, Visualization: Theory and practice in
      science education, Springer Netherlands, 2008.
    
    \bibitem{Nguyen2017}
    H.~Nguyen, D.~A. Case, A.~S. Rose, {NGLview}{\textendash}interactive molecular
      graphics for jupyter notebooks, Bioinformatics 34~(7) (2017) 1241--1242.
    
    \bibitem{seekpath}
    Y.~Hinuma, G.~Pizzi, Y.~Kumagai, F.~Oba, I.~Tanaka, Band structure diagram
      paths based on crystallography, Computational Materials Science 128 (2017)
      140--184.
    
    \bibitem{Clarke2021}
    D.~J. Clarke, M.~Jeon, D.~J. Stein, N.~Moiseyev, E.~Kropiwnicki, C.~Dai,
      Z.~Xie, M.~L. Wojciechowicz, S.~Litz, J.~Hom, J.~E. Evangelista, L.~Goldman,
      S.~Zhang, C.~Yoon, T.~Ahamed, S.~Bhuiyan, M.~Cheng, J.~Karam, K.~M. Jagodnik,
      I.~Shu, A.~Lachmann, S.~Ayling, S.~L. Jenkins, A.~Ma'ayan, Appyters: Turning
      jupyter notebooks into data-driven web apps, Patterns 2~(3) (2021) 100213.
    
    \bibitem{Samuel2018}
    S.~Lau, J.~Hug, nbinteract: generate interactive web pages from jupyter
      notebooks, Master's thesis, EECS Department, University of California,
      Berkeley (2018).
    
    \bibitem{Bokeh}
    {Bokeh Development Team}, Bokeh: Python library for interactive visualization,
      \url{https://bokeh.pydata.org/en/latest} (2018).
    
    \bibitem{AppModeURL}
    {Ole Sch\"utt}, appmode, \url{https://github.com/oschuett/appmode} (2022).
    
    \bibitem{nbconvert}
    {Jupyter Development Team}, nbconvert, \url{https://nbconvert.readthedocs.io}
      (2022).
    
    \bibitem{JupyterBook}
    {Jupyter Book Community}, Jupyter book, \url{https://jupyterbook.org} (2022).
    
    \bibitem{Kim2021}
    B.~Kim, G.~Henke, Easy-to-use cloud computing for teaching data science,
      Journal of Statistics and Data Science Education 29~(sup1) (2021) S103--S111.
    
    \bibitem{DockerURL}
    {Docker, Inc.}, Docker, \url{https://www.docker.com} (2021).
    
    \bibitem{Dokku}
    {Dokku Development Team}, Dokku: The smallest paas implementation you've ever
      seen, \url{https://dokku.com} (2021).
    
    \bibitem{Talirz2020}
    L.~Talirz, S.~Kumbhar, E.~Passaro, A.~V. Yakutovich, V.~Granata, F.~Gargiulo,
      M.~Borelli, M.~Uhrin, S.~P. Huber, S.~Zoupanos, C.~S. Adorf, C.~W. Andersen,
      O.~Sch{\"u}tt, C.~A. Pignedoli, D.~Passerone, J.~VandeVondele, T.~C.
      Schulthess, B.~Smit, G.~Pizzi, N.~Marzari, Materials cloud, a platform for
      open computational science, Scientific Data 7~(1) (2020) 299.
    
    \bibitem{googlecolab}
    {Google LLC}, {Google Colab}, \url{https://colab.research.google.com} (2022).
    
    \bibitem{frenkel2001understanding}
    D.~Frenkel, B.~Smit, Understanding molecular simulation: from algorithms to
      applications, Vol.~1, Elsevier, San Diego, 2001.
    
    \bibitem{teukolsky1992numerical}
    W.~H. Press, B.~P. Flannery, S.~A. Teukolsky, W.~T. Vetterling, Numerical
      recipes in C, The Art of Scientific Computing, Second Edition, Cambridge
      University Press, Cambridge, 1992.
    
    \bibitem{Ainsworth2014}
    S.~Ainsworth, The multiple representation principle in multimedia learning, in:
      R.~Mayer (Ed.), The Cambridge Handbook of Multimedia Learning, Cambridge
      University Press, 2014, pp. 464--486.
    
    \bibitem{izaac2018computational}
    J.~Izaac, J.~Wang, Computational quantum mechanics, Springer, 2018.
    
    \bibitem{shankar2012principles}
    R.~Shankar, Principles of quantum mechanics, Springer Science \& Business
      Media, 2012.
    
    \bibitem{tannor2007introduction}
    D.~J. Tannor, Introduction to quantum mechanics: a time-dependent perspective,
      University Science Books, Sausalito, California, 2007.
    
    \bibitem{thijssen2007computational}
    J.~Thijssen, Computational Physics, Cambridge University Press, Delft, 2007.
    
    \bibitem{fleck1976time}
    J.~A. Fleck, J.~Morris, M.~Feit, Time-dependent propagation of high energy
      laser beams through the atmosphere, Applied physics 10~(2) (1976) 129--160.
    
    \bibitem{Crouch2001}
    C.~H. Crouch, E.~Mazur, Peer instruction: Ten years of experience and results,
      American Journal of Physics 69~(9) (2001) 970--977.
    
    \bibitem{Harris2020}
    C.~R. Harris, K.~J. Millman, S.~J. van~der Walt, R.~Gommers, P.~Virtanen,
      D.~Cournapeau, E.~Wieser, J.~Taylor, S.~Berg, N.~J. Smith, R.~Kern, M.~Picus,
      S.~Hoyer, M.~H. van Kerkwijk, M.~Brett, A.~Haldane, J.~F. del R{\'i}o,
      M.~Wiebe, P.~Peterson, P.~G{\'e}rard-Marchant, K.~Sheppard, T.~Reddy,
      W.~Weckesser, H.~Abbasi, C.~Gohlke, T.~E. Oliphant, Array programming with
      numpy, Nature 585~(7825) (2020) 357--362.
    
    \bibitem{Virtanen2020}
    P.~Virtanen, R.~Gommers, T.~E. Oliphant, M.~Haberland, T.~Reddy, D.~Cournapeau,
      E.~Burovski, P.~Peterson, W.~Weckesser, J.~Bright, S.~J. van~der Walt,
      M.~Brett, J.~Wilson, K.~J. Millman, N.~Mayorov, A.~R.~J. Nelson, E.~Jones,
      R.~Kern, E.~Larson, C.~J. Carey, {\.{I}}.~Polat, Y.~Feng, E.~W. Moore,
      J.~VanderPlas, D.~Laxalde, J.~Perktold, R.~Cimrman, I.~Henriksen, E.~A.
      Quintero, C.~R. Harris, A.~M. Archibald, A.~H. Ribeiro, F.~Pedregosa, P.~van
      Mulbregt, A.~Vijaykumar, A.~P. Bardelli, A.~Rothberg, A.~Hilboll,
      A.~Kloeckner, A.~Scopatz, A.~Lee, A.~Rokem, C.~N. Woods, C.~Fulton,
      C.~Masson, C.~H{\"a}ggstr{\"o}m, C.~Fitzgerald, D.~A. Nicholson, D.~R. Hagen,
      D.~V. Pasechnik, E.~Olivetti, E.~Martin, E.~Wieser, F.~Silva, F.~Lenders,
      F.~Wilhelm, G.~Young, G.~A. Price, G.-L. Ingold, G.~E. Allen, G.~R. Lee,
      H.~Audren, I.~Probst, J.~P. Dietrich, J.~Silterra, J.~T. Webber,
      J.~Slavi{\v{c}}, J.~Nothman, J.~Buchner, J.~Kulick, J.~L. Sch{\"o}nberger,
      J.~V. de~Miranda~Cardoso, J.~Reimer, J.~Harrington, J.~L.~C. Rodr{\'i}guez,
      J.~Nunez-Iglesias, J.~Kuczynski, K.~Tritz, M.~Thoma, M.~Newville,
      M.~K{\"u}mmerer, M.~Bolingbroke, M.~Tartre, M.~Pak, N.~J. Smith, N.~Nowaczyk,
      N.~Shebanov, O.~Pavlyk, P.~A. Brodtkorb, P.~Lee, R.~T. McGibbon,
      R.~Feldbauer, S.~Lewis, S.~Tygier, S.~Sievert, S.~Vigna, S.~Peterson,
      S.~More, T.~Pudlik, T.~Oshima, T.~J. Pingel, T.~P. Robitaille, T.~Spura,
      T.~R. Jones, T.~Cera, T.~Leslie, T.~Zito, T.~Krauss, U.~Upadhyay, Y.~O.
      Halchenko, Y.~V{\'a}zquez-Baeza, S.~1.0~Contributors, Scipy 1.0: fundamental
      algorithms for scientific computing in python, Nature Methods 17~(3) (2020)
      261--272.
    
    \bibitem{Lam2015}
    S.~K. Lam, A.~Pitrou, S.~Seibert, Numba: a llvm-based python jit compiler,
      Proceedings of the Second Workshop on the LLVM Compiler Infrastructure in HPC
      (2015).
    
    \bibitem{Behnel2011}
    S.~Behnel, R.~Bradshaw, C.~Citro, L.~Dalcin, D.~S. Seljebotn, K.~Smith, Cython:
      The best of both worlds, Computing in Science Engineering 13~(2) (2011)
      31--39.
    
    \bibitem{f2py}
    {NumPy Developers}, F2py user guide and reference manual,
      \url{https://numpy.org/doc/stable/f2py} (2021).
    
    \bibitem{Hunter2007}
    J.~D. Hunter, Matplotlib: A 2d graphics environment, Computing in Science \&
      Engineering 9~(3) (2007) 90--95.
    
    \bibitem{Jon2018}
    {J}on {M}ease, {B}ringing ipywidgets {S}upport to plotly.py, in: {F}atih
      {A}kici, {D}avid {L}ippa, {D}illon {N}iederhut, M.~{P}acer (Eds.),
      {P}roceedings of the 17th {P}ython in {S}cience {C}onference, 2018, pp. 69 --
      76.
    
    \bibitem{bqplotURL}
    {bqplot Development Team}, bqplot, \url{https://github.com/bqplot/bqplot}
      (2022).
    
    \bibitem{Ahrens2005}
    J.~AHRENS, B.~GEVECI, C.~LAW, 36 - paraview: An end-user tool for large-data
      visualization, in: C.~D. Hansen, C.~R. Johnson (Eds.), Visualization
      Handbook, Butterworth-Heinemann, Burlington, 2005, pp. 717--731.
    
    \bibitem{Traitlets}
    {IPython Development Team}, Traitlets, \url{https://traitlets.readthedocs.io}
      (2015).
    
    \bibitem{Patwardhan2017}
    M.~Patwardhan, S.~Murthy, Designing reciprocative dynamic linking to improve
      learners' representational competence in interactive learning environments,
      Research and Practice in Technology Enhanced Learning 12 (2017) 10.
    
    \end{thebibliography}
\end{document}